\documentclass[]{raa}
\usepackage{graphicx,times}
\usepackage{natbib}
\usepackage{lscape}

\providecommand{\nus}{$\nu^{\rm S}_{\rm peak}$}
\providecommand{\nuc}{$\nu^{\rm EC}_{\rm peak}$}
\providecommand{\nufs}{$\nu^{\rm S}_{\rm peak}F(\nu^{\rm S}_{\rm peak})$}
\providecommand{\nufc}{$\nu^{\rm EC}_{\rm peak}F(\nu^{\rm EC}_{\rm peak})$}

\providecommand{\nue}{$\nu_{\rm ext}$}

\begin{document}

   \title{On the external photon fields in Fermi bright blazars
}

 \volnopage{ {\bf 2011} Vol.\ {\bf  } No. {\bf XX}, 000--000}
   \setcounter{page}{1}

   \author{Xiaoyan Gao\inst{1,2,3}
   \and Jiancheng Wang\inst{1,2}
   \and Ming Zhou\inst{1,2,3}
   }

   \institute{National Astronomical Observatories/Yunnan Observatories,
   Chinese Academy of Sciences, kunming, Yunnan province
    650011, P.R. China {\it xy-gao@ynao.ac.cn}\\
    \and
    Key Laboratory for the Structure and Evolution of Celestial Objects,
    Chinese Academy of Sciences, kunming, Yunnan province
    650011, P.R. China\\
    \and
    Graduate School, Chinese Academy of Sciences, Beijing, P.R. China
\vs \no
   {\small Received [year] [month] [day]; accepted [year] [month] [day] }
}

\abstract{The external Compton (EC) model is used to study the high energy emission
of some blazars, in which the external photon field is considered to dominate inverse Compton
radiation. We explore the properties of external photon field through analyzing the \textsl{FERMI}
LAT bright AGN sample within three months detection. In the sample, assuming the high energy
radiation of low synchrotron peaked blazars from the EC process, we find that the external
photon parameter $U_{\rm ext}/\nu_{\rm ext}$ may not be a constant. Calculating synchrotron and inverse Compton (IC) luminosity from the
quasi-simultaneous broadband spectrum energy distributions (SEDs), we find that they have
an approximately linear relation. This indicates that the ratio of external photon and
magnetic energy density is a constant in the comoving frame, implying that the Lorentz
factor of the emitting blob depends on external photon field and magnetic field. The result
gives a strong constraint on the jet dynamic model.
}

\authorrunning{X.-Y. Gao \& J.-C. Wang }            
\titlerunning{On the external photon fields in Fermi bright blazars}  
\maketitle


%
%
\section{Introduction}           
\label{sect:intro}
Blazars, presenting weak (or even absent) broad emission
line and large optical polarization, are radio loud subclass of active galactic
nuclei (AGNs). They are thought to be objects emitting non-thermal radiation
across the entire electromagnetic spectrum from a relativistic jet which is
closely along the line of sight and hence appears Dopplor boosted (Begelman \&
Rees \cite{begelmanr}; Urry \cite{urry}). The broadband spectrum energy
distribution of blazars are double bumped with a first peak usually
in the soft-to-medium X-ray range, and a second one in GeV-TeV energies band
(Sambruna et. al.\cite{sambruna}). The first peak is generally dominated
by synchrotron emission (Marscher \& Gear \cite{marscher}), and the second one is uncertain.

The high energy emission is usually attributed by Compton upscattering of
soft photons. The soft photons may come from synchrotron process called synchrotron
self Compton (SSC) model (Maraschi et al.\cite{maraschi}; Marscher \& Gear\cite{marscher}),
or from external field photons which is called External Compton (EC) model.
The external field photons can possibly include accretion disk radiation
(e.g., Dermer et. al.\cite{dermer}; Dermer \& Schlickeiser\cite{dermarsch}),
broad line region (BLR) (Sikora et al.\cite{sikora}; Blandford \& Levinson\cite{blandfordl};
Ghisellini \&Madau \cite{ghisellinim}; Dermer et al. \cite{dermers}),
dust torus (Bla$\dot{\rm z}$ejowski et al.\cite{blazejowski}) or synchrotron
emission from other regions(Georganopoulos \& Kazanas \cite{georganopoulosk};
Ghisellini \& Tavecchio\cite{ghisellinit}). All these different scenarios
have been tested by specific sources.

Blazars contains BL Lac objects and flat spectrum radio quasars (FSRQs).
In previous literatures, BL Lac objects are often subdivided into two or three subclass
depending on their SED which was first introduced by Padovani \& Giommi \cite{padovanig}.
Depending the peak energy of the synchrotron emission, BL Lacs are classified into
low-frequency or high-frequency BL Lac objects, called LBL and HBL
respectively. Abdo et. al. \cite{abdo} give a new classification to
all types of non-thermal dominated AGNs. They are low synchrotron peaked (LSP),
intermediate synchrotron peaked (ISP) and high synchrotron peaked (HSP) blazars,
defined by the location of the low energy SED peak. The LSP sources, consisting
of flat spectrum radio quasars and low frequency peaked BL Lac objects (LBLs),
have the synchrotron peak in the far-infrared or infrared regime with \nus $\leq 10^{14}$ Hz.
The ISP sources, mainly including LBLs and intermediate BL Lac objects (IBLs),
have their synchrotron peak in optical-UV frequencies with $10^{14}\leq$
\nus$\leq 10^{15}$ Hz, while the HSP objects, almost all known to be high-frequency-peaked
BL Lac objects (HBLs), have their synchrotron peak at X-ray energies with \nus $> 10^{15}$ Hz.

In the framework of leptonic models, the blazar sequence FSRQ$\rightarrow$ LBL
$\rightarrow$ IBL$\rightarrow$ HBL is thought to be a decreasing contribution
of external radiation fields to radiative cooling of electrons and production of
high energy radiation (Ghisellini et al. \cite{ghisellinic}; Celotti \& Ghisellini \cite{celottig}).
The $\gamma$ ray radiation of FSRQs always thought to be attributed by EC model\cite{caob08}.
In this paper we use the sample of LSP blazars from the first three month observation of
$\textit{Fermi Gamma Ray Space Telescope}$ to study the properties of high energy emission and
external photon field based on the EC model. The peak frequencies and fluxes are obtained through
fitting the SEDs which are quasi-simultaneous multifrequency measurement instead of using a
empirical method from spectral slopes $\alpha_{\rm ox}$ and $\alpha_{\rm ro}$. We find that
the external photon parameter in the observer frame $U_{\rm ext}/\nu_{\rm ext}$ is not constant.
We also find that there is a linear correlation between synchrotron luminosity $L_{\rm S}$
and inverse Compton (IC) luminosity $L_{\rm IC}$, implying that the ratio between the external
radiation and the magnetic field energy densities inside the source (in the comoving frame)
depends on Lorentz factor. In section 2 we present the sample of LSP sources. In section 3
we show the statistical method and results. The discussion the results are shown in section 4.

We adopt a concordance cosmology with $H_{\rm0}=70~\rm km~s^{-1}~Mpc^{-1}$,
$\Omega_{\rm \Lambda}=0.7$ and $\Omega_{\rm M}=0.3$.


\section{The sample}
\label{sect:sample}
The Large Area Telescope (LAT) on broad the $\textit{Fermi Gamma Ray Space Telescope}$,
launched on 2008 June 11, provides unprecedented sensitivity in $\gamma-$ray
band (20 MeV to over 300 GeV, Atwood et. al\cite{atwood}). The first three
months, from 2008 August 4 to October 31, of operations in the sky-survey
mode led to the compilation of a list of 106 high confidence $\gamma-$ray
sources. Abdo et. al.\cite{abdo} give the quasi-simultaneous SEDs of 48 blazars
from the LAT observation. They include 23 FSRQs, 19 BL Lacs and one unknown
type blazar, which have certain redshifts.

The sample we adopted consists of 30 low synchrotron peak (LSP)
blazars, including 23 FSRQs and 7 low energy peak BL Lacs. In view of blazar sequence, the gamma-ray emission of
these LSPs could be dominated by external compton radiation. We also
present 8 high energy peak BL Lacs for contrast. Usually the HBLs lack the emission line,
the SSC radiation could dominate their gamma-rays. All sources are listed in
table~\ref{table:1}. Column 1 gives the source name presented by Abdo et al.
\cite{abdo09}; Column 2 gives the redshift; Column 3, 4, 5 and 6 are the peak
frequencies and fluxes of synchrotron and inverse Compton (IC) emission. We
use the peak frequencies \nus, \nuc, and the peak fluxes \nufs, \nufc, given by Abdo et. al.\cite{abdo}
who using a simple third-degree polynomial to fit the SED (Kubo et al. \cite{kubo}), where the
data are quasi-simultaneous multifrequency observation. Column 7 lists the total synchrotron
luminosity; Column 8 is the total IC luminosity; Column 9 is the IC luminosity calculated by
the EC model; Column 10 gives the classifications.


\section{Model and Test}
\label{sect:test}

We consider the homogeneous EC model to study the properties of high energy SED for
LSP sources. The synchrotron and IC peak frequencies are given by
(Tavecchio et. al.\cite{tavecchiom})
\begin{equation}
   \nu^{\rm S}_{\rm peak}=\frac{4}{3}\nu_{\rm L}\gamma_{\rm b}^{2}\delta,
\end{equation}
and
\begin{equation}
   \nu^{\rm EC}_{\rm peak}=\frac{4}{3}\nu_{\rm ext}\gamma_{\rm b}^{2}\Gamma\delta,
\end{equation}
where $\nu_{\rm L}=eB/(2\pi m_{\rm e}c)$ is the Larmor frequency, $\gamma_{\rm b}$
is the Lorentz factor of electron spectral break,
and $\nu_{\rm ext}$ is the concentrated frequency of external photons. $\Gamma$ is the
Lorentz factor of the emitting blob. While $\delta$ is the Doppler factor defined by
$\delta=1/(1-v/c \rm cos \theta)$, $v$ is the relativistic bulk velocity, and
$\theta$ is the angle of the jet relative to observer's line of sight.

The ratio between the total luminosity of synchrotron and IC radiation is related
to the ratio between external field and magnetic energy density as
\begin{equation}
   \frac{L_{\rm IC}}{L_{\rm S}}=\frac{U^{'}_{\rm ext}}{U^{'}_{\rm B}}\simeq\frac{17}
   {12}\frac{\Gamma^{2}U_{\rm ext}}{U^{'}_{\rm B}},
\end{equation}
where $U^{'}_{\rm ext}\simeq(17/12)\Gamma^{2}U_{\rm ext}$ is the external photon
energy density in the comoving frame (Ghisellini \& Madau \cite{ghisellinim}), and
$U^{'}_{\rm B}$ is the magnetic field energy densities inside the source (in the comoving frame).
Here we declare that parameters with prime (`` $\rm '$ '') presents in coming  frame,
otherwise are in observer frame.

From the above equations we obtain
\begin{equation}
   \frac{L_{\rm IC}}{L_{\rm S}}\simeq\frac{17e^2}{6\pi m^2_{\rm e}c^2}
   \frac{U_{\rm ext}}{\nu^{2}_{\rm ext}}\left(\frac{\nu^{\rm EC}_{\rm peak}}
   {\nu^{\rm S}_{\rm peak}}\right)^{2}.
\end{equation}

\begin{figure}
    \centering
     \includegraphics[width=13cm]{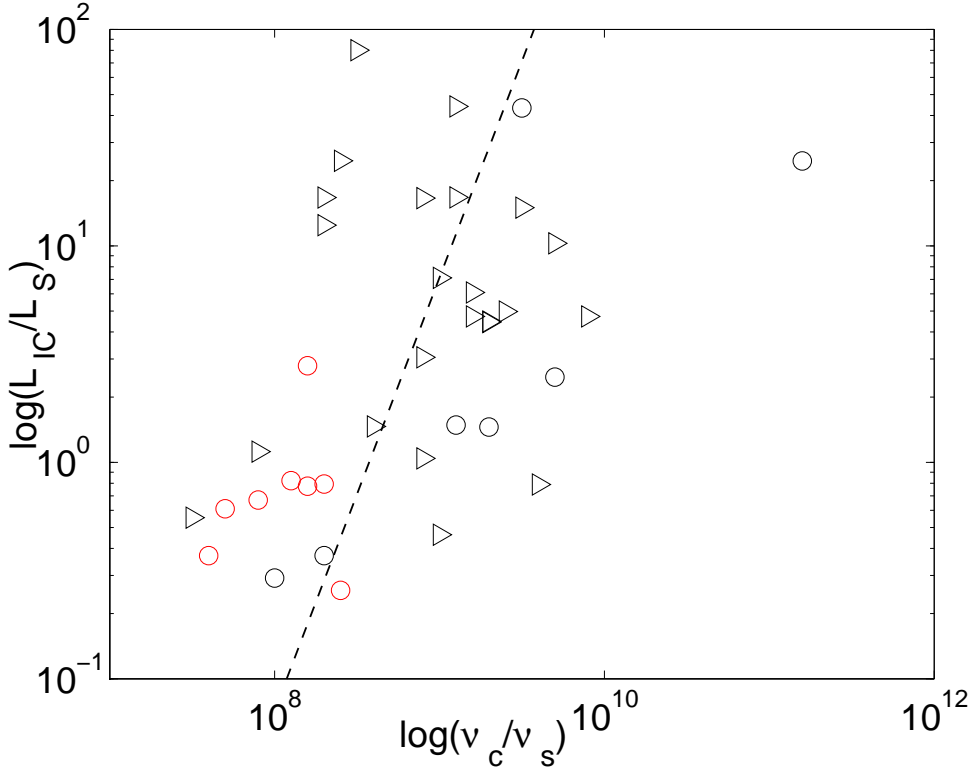}
      \caption{The relation between $L_{\rm IC}/L_{\rm S}$ and \nuc/\nus. Triangles: FSRQs (LSPs);
      black circles: BL Lac objects (LSPs) and red circles: BL Lacs which are HSPs. The dashed line is the fitted line with the slope of 2.}
   \label{Fig1}
\end{figure}

The SEDs of the sources in the sample are fitted by Abdo et al. \cite{abdo}. We use the SEDs to
obtain the differential luminosity of synchrotron and IC radiation. We then calculate the total
luminosity $L_{\rm S}$ and $L_{\rm IC}$ by integrating the differential luminosity. The results
are shown in column (7) and column (8) of table~\ref{table:1}. Assuming $L_{\rm IC}$ from external
Compton emission mainly, we examine the relation of $L_{\rm IC}/L_{\rm S}$ and
$\nu^{\rm EC}_{\rm peak}/{\nu^{\rm S}_{\rm peak}}$ shown in Figure ~\ref{Fig1}, and find that the relation is rather disperse.
No obvious correlation implies that the parameter
${U_{\rm ext}}/{\nu^{2}_{\rm ext}}$ of the external photons is not a constant from equation (4).
The result is different with common scenario on the external photon from the BLR (Ghisellini et al.
\cite{ghisellinic}; Celotti \& Ghisellini \cite{celottig}; Ghisellini \& Madau
\cite{ghisellinim}; Tavecchio \& Ghisellini\cite{tavecchiog}). In particular, we take the common
values of $U_{\rm BLR}=2.65 \times10^{-2}~\rm erg~cm^{-3}$ and \nue=$2\times10^{15} \Gamma\rm~Hz$,
and calculate the EC luminosity $L_{\rm EC}$ from $L_{\rm S}$ using the equation (4) shown in column (9).
We find that the $L_{\rm EC}$ of many sources is larger than the $L_{\rm IC}$ from the SEDs, indicating
that the previous results on the external photon field have questionable validity.

\begin{figure}
 \centering
     \includegraphics[width=13cm]{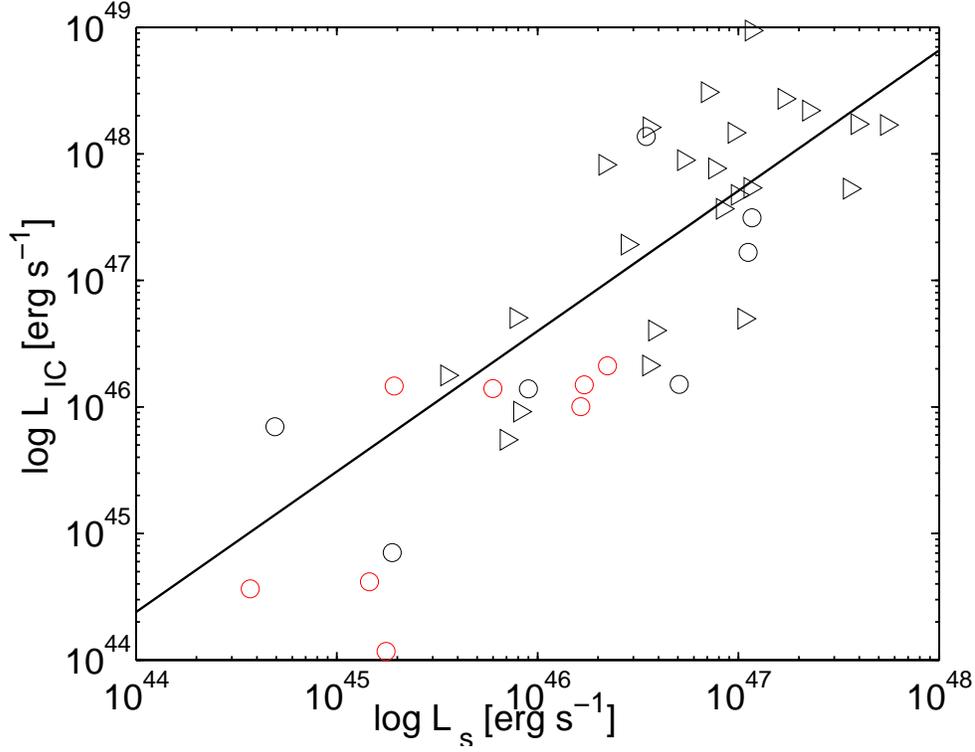}
      \caption{The relation of the total luminosity $L_{\rm S}$ and $L_{\rm IC}$.
      Symbols are the same as in figure ~\ref{Fig1}.}
         \label{Fig2}
\end{figure}

We also show the relation between $L_{\rm IC}$ and $L_{\rm S}$ in Figure~\ref{Fig2}, indicating
a significant correlation. The best fit to the $L_{\rm IC}$ -- $L_{\rm S}$ relationship is
\begin{equation}
\rm LogL_{\rm IC}=1.1 \rm LogL_{\rm S}-4.5
\end{equation}
with a correlation coefficient $R=0.58$, and the relationship is nearly linear, spanning a
wider range of both luminosity. The relation indicates that $L_{\rm IC}/L_{\rm S}$ is an
approximate constant, implying the ratio of external photon and magnetic energy density to
be constant in the comoving frame. If the gamma-ray blob is inside the BRL, from equation (3),
we will obtain $\Gamma\propto (U^{'}_{\rm B}/U^{'}_{\rm ext})^{1/2}$,
implying that the motion of the jet could relate to the magnetic field and external photon.
Usually the magnetic field takes an important role in the acceleration and collimation of the
relativistic jet (Leismann et al. \cite{lei05}, McKinney \cite{mck06}, Keppens et al. \cite{kep08}),
while the external photons can affect the jet dynamics through radiative drag (Sikora et. al. \cite{sik96},
Luo \& Protheroe \cite{luo99}). Therefore, the relation we obtained provides an observational
evidence for the balance between the acceleration and deceleration of the jet. The main contribution
for high energy radiation of HSPs is always considered to be the SSC emission, and the EC emission
could be a supplement for high energy radiation. For comparison, we plot HSPs in Figure ~\ref{Fig2} and
find that HSPs behave similarly with LSPs, locating at low luminosity range.

\section{Conclusion and Discussion}
\label{sect:conclusion}
The BL Lac objects in our sample include two subsamples: LBLs and HBLs.
The LBLs together with FSRQs is considered as LSPs (Abdo et. al.\cite{abdo}) as mentioned above.
External Compton scattering is likely to be the main emission process responsible
for the copious amounts of $\gamma$-rays produced in LSP blazars. The soft seed photons
outside the $\gamma$-ray emitting region dominate gamma-ray emission.
Usually the HBLs lack the broad emission line, the SSC could dominate their gamma-rays. The equation (3)
is obviously invalid for the HBLs, while the HSPs in the figure 1 only show their position
for comparison with other sources and don't present their relations with gamma-ray emission mechanism.
For LBLs, if the gamma-rays originate from the blob inside the BLR and the size of the BLR is proportional
to $L^{0.5}$ (Celotti \& Ghisellini\cite{celottig}), the energy density
and peak frequency of external photons will be a constant with good approximation. It is indicated that
the ratio between EC and synchrotron luminosity just depends on the peak frequency
(\nus/\nuc)$^2$. In the paper we have presented the statistic analysis of a sample of 30 LSP
blazars from the $\textit{Fermi}$-LAT observation. After examining the relation between
$L_{\rm IC}/L_{\rm S}$ and \nuc/\nus for LSP blazars, we find there is almost no correlation between them.
It means that the parameter ${U_{\rm ext}}/{\nu^{2}_{\rm ext}}$ of the external photons is not a constant or
the equation (3) is invalid for some LSPs due to the gamma-ray blob locating outside the BLR.
Our statistical results show that the gamma-rays of LSPs could be produced at different external photon fields.
There are two kinds of emissive scenarios. (1) The gamma-ray blob is
inside the BLR, the figure 1 implies that the external photons could come from many components
including the BLR and have not constant energy density\cite{bail09,caob08}.
(2) The gamma-ray blob is outside the BLR, the external photons from the BLR is
Doppler deboosted for the blob. The relation given by equation (4) is invalid,
leading no correlation between $L_{\rm IC}/L_{\rm S}$ and \nuc/\nus,
shown in figure 1. We also consider the relation of $L_{\rm IC}$ and $L_{\rm S}$, and find
an approximate linear relation between them, although the statistical obviousness is not high due to
the lack of large sample. This result shows that
the ratio of external field and magnetic energy density in the emitting blob is an approximate constant.
This implies that the Lorentz factor depends on the external photon field and magnetic field when the blob is inside the BLR.
The result indicates an evidence of balancing the jet motion under the action of radiative drag
and magnetic derive, and can be used to constrain the jet dynamic model.

\normalem
\begin{acknowledgements}
We acknowledge the financial supports from the National Natural
Science Foundation of China 10778702, the National
Basic Research Program of China (973 Program 2009CB824800), and the
Policy Research Program of Chinese Academy of Sciences (KJCX2-YW-T24).
\end{acknowledgements}

\begin{landscape}
    \begin{table*}
     \caption{List of sources and basic parameters studied in this paper.
      (1) source LAT names; (2) redshift; (3) synchrotron peak frequency in unit of [Hz];
      (4) synchrotron peak flux in unit of [$\rm erg~cm^{-2}~s^{-1}$]; (5) IC peak
       frequency in unit of [Hz]; (6) IC peak flux in unit of [$\rm erg~cm^{-2}~s^{-1}$];
       (7) synchrotron luminosity with [$\rm erg s^{-1}$]; (8) IC luminosity in unit of [$\rm erg s^{-1}$];
       (9) IC luminosity calculated by the EC model in unit of [$\rm erg s^{-1}$]; (10) classification.}
       \label{table:1}
     \centering
     \begin{tabular}{llllllllll}
     \hline\hline
              $$
            0FGL &  z  & log\nus & log$(\nu F\nu)^{\rm S}_{\rm paek}$ & log$\nu^{\rm IC}_{\rm peak}$ & log$(\nu F\nu)^{\rm IC}_{\rm paek}$ &
             log$L_{\rm S}$ & log$L_{\rm IC}$ &  log$L_{\rm EC}$ & Class.\\
            (1) & (2) & (3) & (4) & (5) & (6) & (7) & (8)& (9)& (10)  \\
            \hline
            J0137.1+4751 & 0.859 & 13.6 & -10.8 & 22.6 & -10.6 &   47.03 & 46.70 & 47.29 & FSRQ\\
            J0210.8-5100 & 1.003 & 12.5 & -10.5 & 22.4 & -10.2 &   47.06 & 47.73 & 49.13 & FSRQ\\
            J0229.5-3640 & 2.115 & 13.5 & -10.7 & 21.8 & -10.4 &   46.73 & 47.95 & 45.59 & FSRQ\\
            J0238.4+2855 & 1.213 & 12.8 & -11.7 & 22.1 & -10.8 &   46.92 & 47.57 & 47.79 & FSRQ\\
            J0238.6+1636 & 0.94  & 13.5 & -10.0 & 23.2 & -9.90 &   47.07 & 47.50 & 48.73 & BL Lac\\
            J0349.8-2102 & 2.944 & 12.9 & -10.7 & 21.8 & -10.2 &   47.35 & 48.34 & 47.42 & FSRQ\\
            J0423.1-0112 & 0.916 & 13.4 & -11.3 & 21.7 & -10.3 &   46.34 & 47.91 & 45.20 & FSRQ\\
            J0428.7-3755 & 1.03  & 13.3 & -11.0 & 22.8 & -10.2 &   46.54 & 48.14 & 47.81 & BL Lac\\
            J0457.1-2325 & 1.003 & 13.1 & -10.9 & 22.8 & -9.90 &   46.88 & 47.89 & 48.55 & FSRQ\\
            J0531.0+1331 & 2.07  & 12.8 & -11.0 & 21.3 & -9.80 &   47.06 & 48.98 & 46.33 & FSRQ\\
            J0538.8-4403 & 0.892 & 13.4 & -10.6 & 22.7 & -10.1 &   47.05 & 47.22 & 47.92 & BL Lac\\
            J0730.4-1142 & 1.598 & 13.1 & -11.1 & 22.6 & -10.0 &   46.98 & 48.17 & 48.25 & FSRQ\\
            J0855.4+2009 & 0.306 & 13.4 & -9.8  & 21.4 & -10.5 &   46.71 & 46.18 & 44.97 & BL Lac\\
            J0921.2+4437 & 2.19  & 13.4 & -10.9 & 22.0 & -10.6 &   47.55 & 47.73 & 47.02 & FSRQ\\
            J1159.2+2912 & 0.729 & 13.1 & -11.2 & 22.0 & -10.5 &   46.58 & 46.60 & 46.65 & FSRQ\\
            J1229.1+0202 & 0.158 & 13.5 & -10.7 & 21.0 & -9.60 &   46.55 & 46.33 & 43.82 & FSRQ\\
            J1256.1-0548 & 0.536 & 12.6 & -9.8  & 22.2 & -10.3 &   45.84 & 45.74 & 47.31 & FSRQ\\
            J1310.6+3220 & 0.997 & 13.1 & -10.3 & 22.5 & -10.4 &   46.99 & 47.68 & 48.06 & FSRQ\\
            J1457.6-3538 & 1.424 & 13.6 & -10.9 & 22.7 & -10.2 &   47.23 & 48.44 & 47.69 & FSRQ\\
            J1504.4+1030 & 1.839 & 13.6 & -11.7 & 22.9 & -9.80 &   45.89 & 46.70 & 46.56 & FSRQ\\
            J1512.7-0905 & 0.36  & 13.1 & -10.9 & 22.3 & -9.70 &   46.56 & 48.21 & 47.02 & FSRQ\\
            J1522.2+3143 & 1.487 & 13.3 & -10.6 & 22.4 & -10.2 &   46.45 & 47.28 & 46.71 & FSRQ\\
            J1719.3+1746 & 0.137 & 13.5 & -10.3 & 24.7 & -10.7 &   44.69 & 45.84 & 49.36 & BL Lac\\
            J1751.5+0935 & 0.322 & 13.1 & -10.8 & 22.2 & -10.3 &   45.95 & 46.14 & 46.42 & BL Lac\\
            J1849.4+6706 & 0.657 & 13.5 & -10.6 & 22.5 & -10.5 &   45.91 & 45.96 & 43.98 & FSRQ\\
            J2143.2+1741 & 0.213 & 14.1 & -10.4 & 22.0 & -10.5 &   47.74 & 48.23 & 47.81 & FSRQ\\
            J2202.4+4217 & 0.069 & 13.6 & -10.1 & 21.9 & -10.8 &   45.28 & 44.85 & 44.14 & BL Lac\\
            J2254.0+1609 & 0.859 & 13.6 & -11.5 & 22.5 & -9.80 &   45.55 & 46.25 & 46.21 & FSRQ\\
            J2327.3+0947 & 1.843 & 13.1 & -11.0 & 21.5 & -10.3 &   47.59 & 48.24 & 48.46 & FSRQ\\
            J2345.5-1559 & 0.621 & 13.3 & -9.5  & 22.5 & -10.7 &   46.85 & 48.49 & 45.91 & FSRQ\\
            J0033.6-1921 & 0.61  & 16.1	& -11.1	& 24.3 & -11.1 &   46.23 & 46.17 & 44.90 & BL Lac\\
            J0449.7-4348 & 0.205 & 15.6	& -10.2	& 23.9 & -10.5 &   46.29 & 46.17 & 44.15 & BL Lac\\
            J0507.9+6739 & 0.416 & 16.6	& -10.7	& 24.3 & -10.5 &   46.35 & 46.32 & 44.01 & BL Lac\\
            J1015.2+4927 & 0.2   & 16.3	& -10.5	& 24.5 & -10.6 &   45.78 & 46.15 & 44.44 & BL Lac\\
            J1104.5+3811 & 0.03  & 16.6	& -9.4	& 25   & -9.9  &   45.16 & 44.62 & 44.23 & BL Lac\\
            J1653.9+3946 & 0.033 & 17.1	& -10.3	& 24.7 & -10.5 &   45.25 & 44.07 & 42.71 & BL Lac\\
            J2000.2+6506 & 0.047 & 16.6	& -10	& 24.7 & -10.5 &   44.57 & 44.56 & 43.03 & BL Lac\\
            J2158.8-3014 & 0.116 & 16   & -9.7	& 23.9 & -10.2 &   46.22 & 46.00 & 44.28 & BL Lac\\
            \hline
       $$
       \end{tabular}
 \end{table*}
\end{landscape}
\label{lastpage}
\end{document}